\documentclass[%
 reprint,
superscriptaddress,
 amsmath,amssymb,
 aps,
]{revtex4-2}

\usepackage{graphicx}
\usepackage{dcolumn}
\usepackage{bm}
\usepackage{dsfont}
\usepackage{hyperref}
\usepackage{braket}
\usepackage{xcolor}

\usepackage[normalem]{ulem}

\newcommand{\rev}[1]{\textcolor{black}{#1}}
\newcommand{\edit}[1]{\textcolor{black}{#1}}

\begin{document}

\title{Shortcut-to-Adiabatic Controlled-Phase Gate in Rydberg atoms}

\author{Luis S. Yag{\"u}e Bosch}
\affiliation{%
Institute of Theoretical Physics, Heidelberg University, Philosophenweg 12, 69120 Heidelberg, Germany
}
\author{Tim Ehret}
\altaffiliation[now at: ]{Faculty of Physics, Vienna University, Boltzmanngasse 5, 1090 Wien, Austria}
\affiliation{%
Institute of Theoretical Physics, Heidelberg University, Philosophenweg 12, 69120 Heidelberg, Germany
}
\author{Francesco Petiziol}%
  \affiliation{Technische Universit\"at Berlin, Institut f{\"u}r Theoretische Physik, Hardenbergstrasse 36, Berlin 10623, Germany}
\author{Ennio Arimondo}
  \affiliation{Dipartimento di Fisica, Universit{\`a} di Pisa, Largo Pontecorvo 3, 56127 Pisa, Italy}
  \affiliation{Istituto Nazionale di Ottica, Consiglio Nazionale delle Ricerche, Universit{\'a} di Pisa, Largo Pontecorvo 3, 56127 Pisa, Italy}
  \author{Sandro Wimberger}%
  \email{sandromarcel.wimberger@unipr.it}
\affiliation{%
Dipartimento di Scienze Matematiche, Fisiche e Informatiche,
Universit{\`a} di Parma, Parco Area delle Scienze 7/A, 43124 Parma, Italy
}
\affiliation{%
INFN, Sezione di Milano Bicocca, Gruppo Collegato di Parma,
Parco Area delle Scienze 7/A, 43124 Parma, Italy
}

\begin{abstract}
A shortcut-to-adiabatic protocol for the realization of a fast and high-fidelity controlled-phase gate in Rydberg atoms is developed. The adiabatic state transfer, driven in the high-blockade limit, is sped up by compensating nonadiabatic transitions via oscillating fields that mimic a counterdiabatic Hamiltonian. High fidelities are obtained in wide parameter regions. The implementation of the bare effective counterdiabatic field, without original adiabatic pulses, enables to bypass gate errors produced by the accumulation of blockade-dependent dynamical phases, making the protocol efficient also at low blockade values. As an application toward quantum algorithms, how the fidelity of the gate impacts the efficiency of a minimal quantum-error correction circuit is analyzed.
\end{abstract}

\maketitle

\section{Introduction}
\label{sec:intro}

Rydberg atoms \cite{Gallagher2005} have been established as a platform for many applications in quantum physics over the last 30 years, going from old beam experiments \cite{Rempe1987, Varcoe2000, Haroche2001, Haroche2013} to much better controlled cold-atom setups in which single Rydberg atoms are hold in position with optical tweezers \cite{Graham2019, Levine2019, Scholl2021, Ebadi2021}. We are interested in the use of cold Rydberg atoms for quantum information protocols \cite{Morgado2021, Adams2020, Saffman2010, Graham2019}, in view of their ability to implement single well-controlled quantum gates with high fidelity \cite{Saffman2016, Su2016, Levine2019, Graham2019, Pelegri2022, Pagano2022, Jandura2022, Robicheaux2021, Chang2023}.

In particular, adiabatic protocols seem to be promising for large-scale computations because of their intrinsic robustness \cite{Brierley2012, Vitanov2017, Barends2016}. In contrast to sudden or quasi-resonant protocols no precise resonance tuning and pulse timing is required
\cite{Morgado2021, Adams2020, Saffman2010, Saffman2016, Su2016, Levine2019, Graham2019, Pelegri2022, Pagano2022, Jandura2022, Robicheaux2021, Chang2023, Han2016}.
A specific adiabatic protocol for realizing a controlled-phase (CZ) gate has recently been proposed by Saffman et al. \cite{Saffman2020}. An advantage of this adiabatic gate is that it prevents large transient occupation of the Rydberg states, such that the protocol is less exposed to spontaneous emission from such states---a primary source of fidelity loss in this platform. This interacting Rydberg gate works differently from a recent realization of a CZ gate with superconducting qubits based on a different form of interaction \cite{Wang2019}. 

In this work, we explore the possibility to overcome the typical slowness of adiabatic quantum computation protocols by accelerating the CZ gate of \cite{Saffman2020} via shortcuts to adiabaticity~\cite{Torrontegui2013, Odelin2019} (STA). The idea of a shortcut is to find a modified Hamiltonian that realizes the same population transfer with high fidelity but in a nonadiabatic, faster way, and it has been successfully applied to control in few-level systems in multiple experimental platforms, see, e.g.,~\cite{Bason2012, An2016, Zhou2017, Boyers2019, Vepsalainen2019}. In particular, we derive and analyse a control protocol in which nonadiabatic transitions are counteracted via additional oscillating components in the control fields that mimic a counterdiabatic (CD) Hamiltonian~\cite{Petiziol2018, Boyers2019, Claeys2019}.  
\rev{While STA techniques typically address individual state-to-state transfers, we here rather engineer the full propagator by targeting the realization of a complete gate. This more challenging task, which has been traditionally carried out using optimal control \cite{Dorner2005, Treutlein2006, Cozzini2006, Goerz2011, Glaser2015}, was recently also addressed with STA methods~\cite{Martinis2014, Chen2015, Santos2015, Petiziol2019, Shen2019, Yan2019, Petiziol2020, Wang2018}.} 

The focus of the CD protocols is usually on non-interacting systems,
because as a common feature the required CD corrections are hard to be
implemented experimentally, with a few exceptions of integrable many-body systems \cite{Torrontegui2013, Kolo2017, Odelin2019, Hegade2021, Wurtz2022}. This is true, in principle, also for protocols with Rydberg atoms interacting in their excited states, but we demonstrate
that an approximate Hamiltonian not requiring many-body controls nevertheless leads to good final fidelities for a broad range of protocol times and system parameters.

\section{Adiabatic CZ gate} 
\label{sec:adiabaticCZ}

\begin{figure}[t!]
\includegraphics[width=\linewidth]{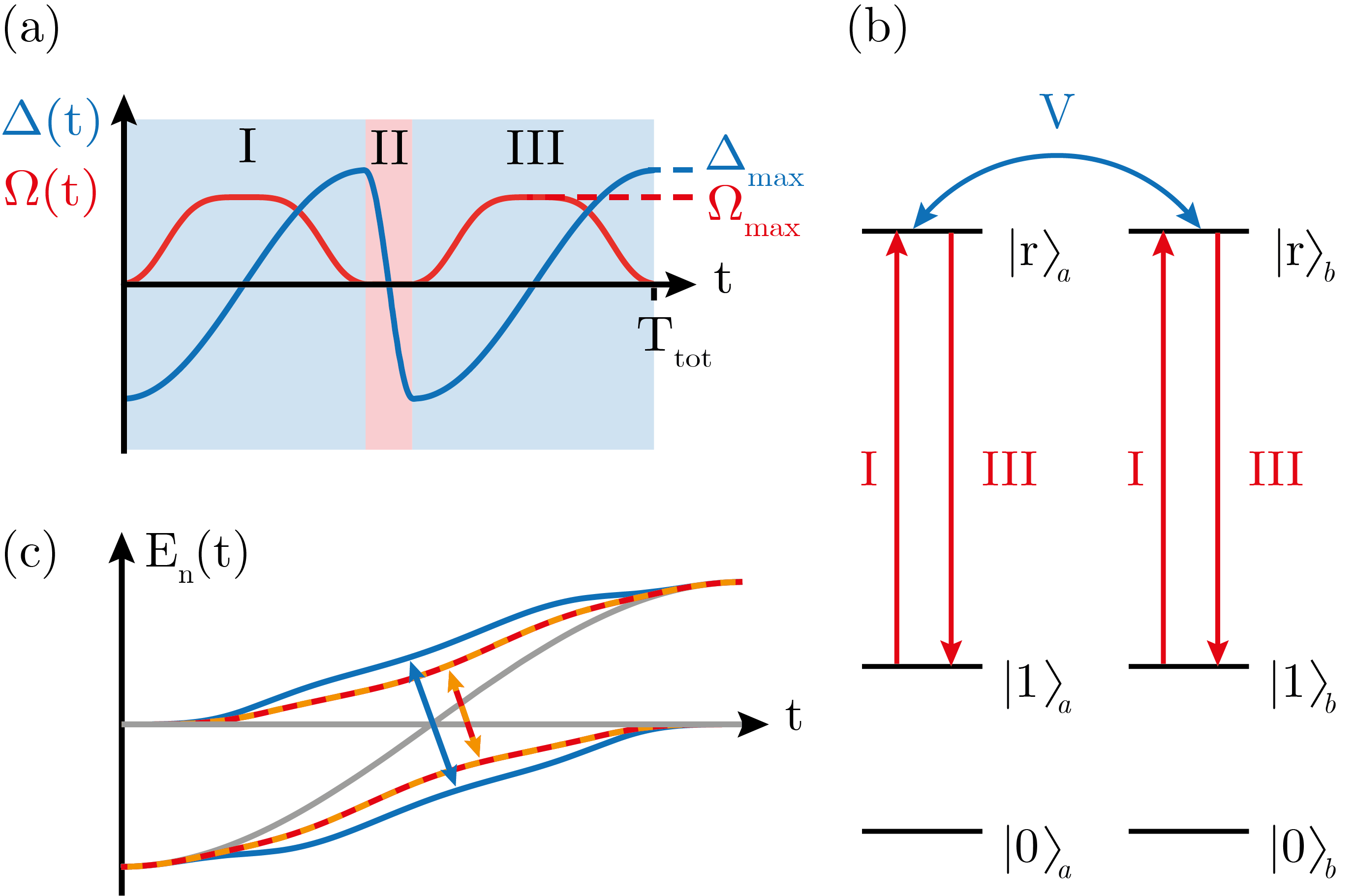}
\caption{(a) Time dependence of the adiabatic pulses. In phase II the detuning is changed continuously. \edit{The explicit form of the curves is given in the text. The peak values of the pulses are $\Omega_\mathrm{max}/2\pi = 17$ MHz and $\Delta_\mathrm{max}/2\pi = 23$ MHz at a total time of $T_\mathrm{tot} = 0.594\;\mu$s. The width is given by $\tau = 0.0945\;\mu$s and the detuning in phase II takes $\approx T_\mathrm{tot}/10$.} (b) Energy levels scheme and sketch of the driving sequence; the atoms are first excited to their Rydberg state (phase I), and are then de-excited (phase III); the adiabatic pulses interpolate the three phases in a continuous manner. (c) Instantaneous energy levels of the driven Hamiltonian of Eq.~\eqref{eq:Ht} as a function of time in phase I; Arrows indicate two avoided crossings between states $\ket{01}$ and $\ket{0r}$ or $\ket{10}$ and $\ket{r0}$, and states $\ket{11}$ and $\ket{1r}+\ket{r1}$, respectively.}
\label{fig:1}
\end{figure}

While the CD approach may be applied to any adiabatic  protocol, we focus our attention on the configuration depicted in Fig.~\ref{fig:1} representing an adiabatic Rydberg driving scheme slightly modified with respect to \cite{Saffman2020}. The two atoms, having \edit{hyperfine levels $\ket{0}$ and $\ket{1}$ of the electronic ground state}, are symmetrically driven with lasers of Rabi frequency $\Omega(t)$ targeting the transition from $\ket{1}$ to a Rydberg state $\ket{r}$, with detuning $\Delta(t)$, and they experience a Rydberg-Rydberg interaction of strength $V$. The time-dependent Hamiltonian, in the frame rotating at the laser frequency, reads
\begin{equation} \label{eq:Ht}
H(t) = H_d(t)\otimes \mathds{1} + \mathds{1}\otimes H_d(t) + V\ket{rr}\!\bra{rr},
\end{equation}
where $H_d(t)$ describes single atom control,
\begin{equation}
H_d(t) = \frac{\Omega(t)}{2} (\ket{r}\!\bra{1} + \ket{1}\bra{r}) + \Delta(t)\ket{r}\!\bra{r}.   
\end{equation}
\edit{Both pulses $\Omega(t)$ and $\Delta(t)$ are represented in Fig.~\ref{fig:1}(a). The Rabi frequency has the following shape typical for adiabatic rapid passage \cite{Saffman2020, Petiziol2020, Morris1963, Vitanov2017}
\begin{equation}
    \Omega(t)=\frac{\Omega_\mathrm{max}}{\mathcal{N}}\left[e^{-(t-t_0)^4/\tau^4}-a-bt(t-2t_0)\right],
\end{equation}
which is centered at $t_0$. The real constants $a$ and $b$ are chosen such that the function and its derivative vanish at the start and end points. $\mathcal{N}$ is a normalization constant such that $\Omega_\mathrm{max}$ is the maximum of the function $\Omega(t)$. The detuning is sine shaped including a continuous sign change in phase II to avoid a sudden jump of the function.} These pulses produce subsequent excitation (phase I) and de-excitation (phase III) of the atoms to state $\ket{r}$, where the Rydberg blockade mechanism sets in, as sketched in Fig.~\ref{fig:1}(b). During phase II, the detuning is smoothly inverted, while the state remains unchanged. The instantaneous energy levels of $H(t)$ as a function of time are shown in Fig.~\ref{fig:1}(c), where two avoided crossings are visible (indicated by arrows), signaling a mixing of states $\ket{01}$ and $\ket{0r}$ (red colour), $\ket{10}$ and $\ket{r0}$ (orange colour) and $\ket{11}$ and $\ket{d_+} = (\ket{1r}+\ket{r1})/\sqrt{2}$ (blue colour), respectively. The pulses are designed for initial states $\ket{01}$, $\ket{10}$ and $\ket{11}$ to accumulate a geometric phase $\pi$ during the de-excitation in the infinite blockade and adiabatic limit. Indeed, in this case the basis states evolve according to 
\begin{eqnarray}
&\ket{1}\to \ket{r}\to -\ket{1}, \\
& \ket{11} \to \ket{d_+} \to -\ket{11}.
\end{eqnarray}
Together with the fact that state $\ket{00}$ has no dynamics induced by the control fields, the phase accumulation produces the CZ gate. 
An advantage of the adiabatic gate is that it drives the atoms almost continuously, which has been observed to be particularly robust against spontaneous emission from the Rydberg state compared to having the atoms sit in state $\ket{r}$~\cite{Saffman2020}.

The goal of the present work is to develop a new driving protocol that produces the same adiabatic evolution as the original protocol, but in a shorter time, or with a higher fidelity in a given total duration. In order to construct such a protocol, we apply the idea of effective counterdiabatic driving (eCD)~\cite{Petiziol2018} based on Floquet-engineering \cite{Eckardt2017, Goldman2014, Bukov2015}: the pulses are modified with the introduction of additional Fourier components, whose amplitude is chosen such that they compensate for nonadiabatic transitions, by mimicking the effect of a counterdiabatic field~\cite{Berry2009, Demirplak2003}. Differently from other STA techniques, which aim at matching the adiabatic dynamics only at the initial and final times~\cite{Demirplak2008, Ibanez2012, Ibanez2013, MartinezGaraot2014, Deffner2014}, the eCD method strives for tracking the adiabatic dynamics closely at all times. This is an advantage for the present protocol since it allows to maintain the low occupation of the Rydberg states guaranteed by the adiabatic sweep. A similar situation is encountered for accelerated Stimulated Raman Adiabatic Passage (STIRAP)~\cite{Petiziol2020, Vepsalainen2019}, where adiabatic following of a dark state protects against emission from the intermediate state. In the following, the states $\ket{00}$ and $(\ket{r1}-\bra{1r})/\sqrt{2}$ will be neglected, since they are not involved in the driven dynamics. Moreover, we consider the regime of large blockade, $V\gg \Omega(t), \Delta(t)$, such that the $\ket{rr}$ state can be adiabatically eliminated (please see Appendix A). In our numerical simulations, all states are included (without adiabatic elimination) to gauge the effect of finite values of $V$.

\section{Accelerated CZ gate}

In order to construct the accelerating protocol, the first step is to determine the exact counterdiabatic field \cite{Berry2009, Demirplak2003, Demirplak2008, Odelin2019} for the Hamiltonian of Eq.~\eqref{eq:Ht}. This is defined as the correction $H_{\mathrm{CD}}(t)$ to the Hamiltonian, such that $H(t) + H_{\mathrm{CD}}(t)$ generates the adiabatic dynamics exactly without restrictions on the total evolution time. Indicating with $E_n(t)$ and $\ket{n(t)}$ the instantaneous eigenvalues and eigenvectors of the adiabatically driven Hamiltonian $H(t)$, this means that an initial eigenstate $\ket{n(0)}$ is mapped by the evolution to the corresponding instantaneous eigenstate $\ket{n(t)}$ for all times, with additional dynamical and geometrical phase factors, 
\begin{equation} \label{eq:Uad}
\exp\left(-i\int_0^t dt' E_n(t') +\int_0^t dt' \braket{n(t')|\partial_{t'} n(t')} \right) \ket{n(t)}.
\end{equation}
When instantaneous energies and eigenvectors are known, and in the absence of degeneracies, the counterdiabatic field can be compactly expressed as~\cite{Berry2009}
\begin{equation} 
H_{\mathrm{CD}}(t) = i \sum_{n, m\ne n} \frac{\ket{n(t)}\bra{n(t)}\partial_t H(t) \ket{m(t)}\bra{m(t)}}{E_m(t) - E_n(t)}.
\end{equation}
We remark that, if $H_{\mathrm{CD}}(t)$ is implemented, it ensures both tracking of the instantaneous eigenstate $\ket{n(t)}$ and fixes the geometric phase factor in Eq.~\eqref{eq:Uad}, while the role of $H(t)$ is to imprint the dynamical phase only. Since the gate is based on the accumulation of only geometric phases, \rev{$H_{\mathrm{CD}}(t)$ would be} sufficient to realise the CZ gate, even in the absence of the original adiabatic driving $H(t)$. This is an interesting property, since in the present protocol, dynamical phases are actually responsible for unwanted phase shifts: the state $\ket{11}$ dynamically accumulates an excess phase \cite{Ehret2022, Maller2015} 
\begin{equation}
\label{eq:phaseerror}
  \varphi = \int_0^{T_\mathrm{tot}} dt\; \Omega^2(t) /4V .  
\end{equation}
This phase is non-negligible for large Rabi frequencies (which would actually be the ideal regime for adiabaticity) and finite $V$. Therefore, by implementing only $H_\mathrm{CD}(t)$, rather than $H(t)$ or $H(t) + H_{\mathrm{CD}}(t)$, one can obtain both a perfect shortcut and avoids such phase errors. 

\begin{figure}[t!]
\includegraphics[width=\linewidth]{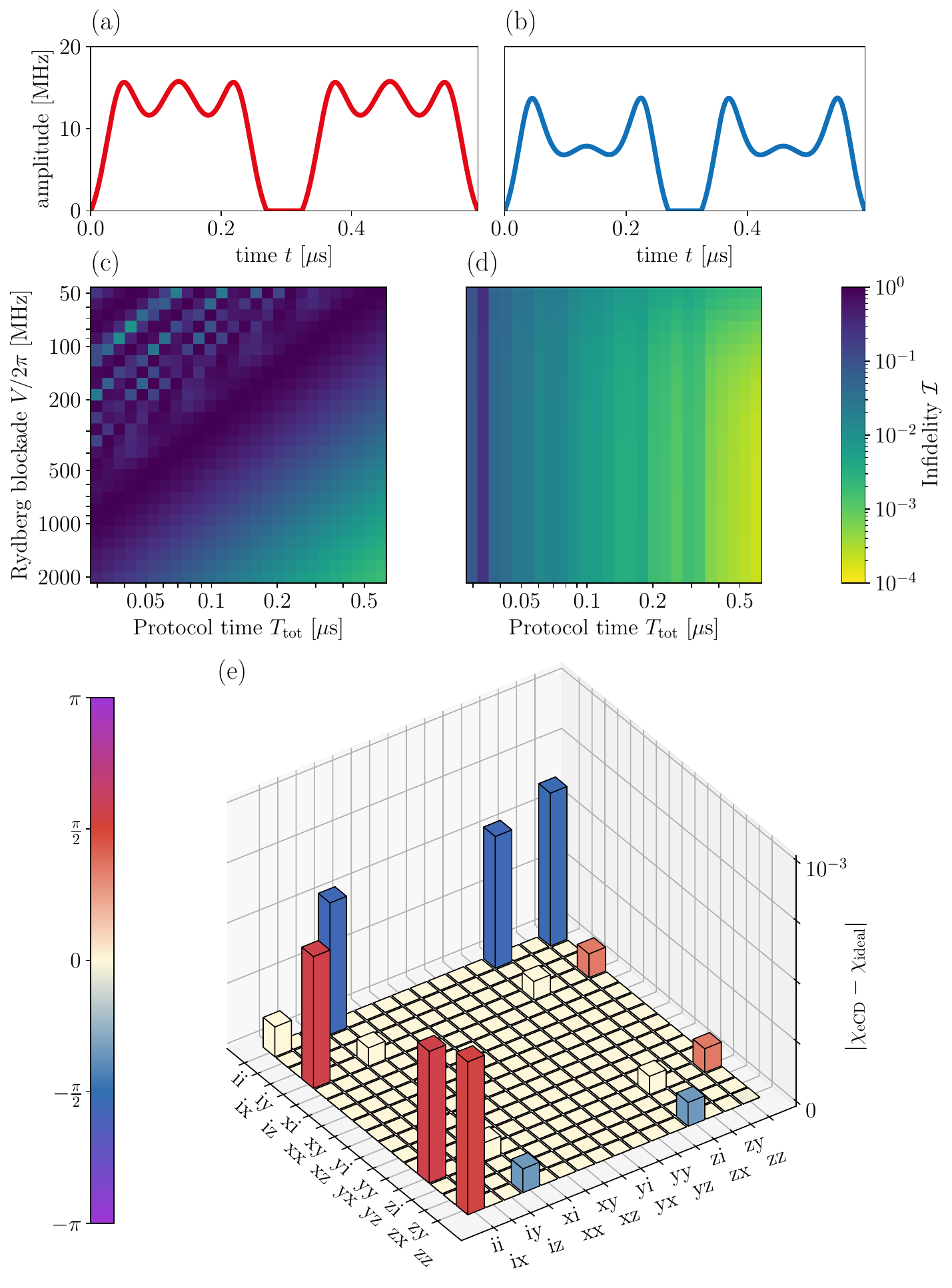}
\caption{Time profile of the amplitudes $f_0(t)$ [(a)] and $f_1(t)$ [(b)] of the counterdiabatic field $H_\text{CD}$. In phase II they vanish as the Rabi pulses are exactly zero. Infidelity for the initial state $(\ket{00}+\ket{11})/\sqrt{2}$ of the adiabatic protocol (c) and of the CD protocol (d) as a function of the total protocol time $T_\mathrm{tot}$ and of the strength of the Rydberg blockade. The change of the detuning in phase II takes $\approx T_\mathrm{tot}/10$.  The pulses have peak values $\Omega_\mathrm{max}/2\pi=17$ MHz and $\Delta_\mathrm{max}/2\pi=23$ MHz at $T_\mathrm{tot}=0.594\;\mu$s as in \cite{Saffman2020} and then scale as $\Omega_\mathrm{max},\;\Delta_\mathrm{max}\propto 1/T_\mathrm{tot}$ in order to maintain adiabatic evolution. The eCD frequency is chosen to be $\omega=300$ MHz. \edit{(e) QPT of the proposed eCD method: difference between the eCD $\chi$-matrix, $\chi_{\mathrm{eCD}}$, and the ideal one. The parameters are the same as in (c) and (d) with fixed $V/2\pi=500$ MHz and $T_\text{tot}=0.594\;\mu$s. The colorbar displays the phase of the matrix elements. The labels $i,x,y,z$ refer to the indices of the operator basis, while the odd (even) axes ticks refer to the upper (lower) row of $\chi$-matrix labels.} }
\label{fig:2}
\end{figure}

To derive $H_{\mathrm{CD}}(t)$ for the Hamiltonian~\eqref{eq:Ht}, we first write $H(t)$ as the sum $H(t)= [H_0(t)\otimes \ket{0}\!\bra{0} + \ket{0}\!\bra{0}\otimes H_0(t)] + H_1(t)$ of two mutually commuting terms corresponding to the subspaces in which at least one of the two atoms is in its $\ket{0}$ ground state (the term in square brackets) or neither is [$H_1(t)$], respectively. Adopting the bases $\{\ket{10},\ket{r0},\ket{01},\ket{0r}\}$ and $\{\ket{11}, \ket{d_+} \}$ for the two separate subspaces, see Appendix A, the Hamiltonians $H_k(t)$ read
\begin{equation} \label{eq:Hk}
H_k(t) = \frac{1}{2}\begin{bmatrix}
0 & \Omega_k(t) \\
\Omega_k(t) & 2\Delta(t)
\end{bmatrix},
\end{equation}
with $\Omega_0(t) = \Omega(t)$ and $\Omega_1(t) = \sqrt{2} \Omega(t)$.
The Hamiltonian $H(t)$ hence describes two uncoupled two-level problems, one defined in each of the \{$\ket{1},\ket{r}$\} single-atom subspaces, and the other associated with the two-qubit subspace $\{\ket{11},\ket{d_+}\}$. These two two-level problems share the same detuning $\Delta(t)$ but the latter has a Rabi frequency which is enhanced by a factor of $\sqrt{2}$, an effect due to the Rydberg blockade~\cite{Saffman2020, Han2016, Levine2019}. Following this observation, the counterdiabatic field $H_{\mathrm{CD}}(t)$ can then readily be expressed using known results for two-level systems~\cite{Berry2009, Demirplak2003}, and, for real-valued $\Omega_k(t)$ and $\Delta(t)$, it involves purely imaginary off-diagonal couplings. It reads $H_{\mathrm{CD}}(t)=H_{\mathrm{CD},0}(t) + H_{\mathrm{CD},1}(t) $, with
\begin{align}
H_{\mathrm{CD},0}(t) = &-\frac{f_0(t)}{2i}\big[\ket{1}\!\bra{r}\otimes\ket{0}\!\bra{0} \nonumber \\
&\quad + \ket{0}\!\bra{0} \otimes \ket{1}\!\bra{r}\big] + \mathrm{H.c.}\ ,\nonumber \\
H_{\mathrm{CD}, 1}(t)  = & - \frac{f_1(t)}{2i} \ket{11}\!\bra{d_+} + \mathrm{H.c.}\ , \label{eq:Hcd01}
\end{align}
where the time-dependent amplitudes $f_k(t)$, $k\in \{0,1\}$, are given by
\begin{equation} \label{eq:fk}
f_k(t) = \frac{\Delta(t)\partial_t\Omega_k(t) - \Omega_k(t)\partial_t {\Delta(t)}}{\Delta^2(t) + \Omega_k^2(t)} .
\end{equation}
The time profile of the these functions is shown in Fig.~\ref{fig:2}(a,b) for the adiabatic pulse shapes considered here.
A direct implementation of the CD field would be experimentally challenging for two reasons. First, it requires the introduction of additional pulses with imaginary Rabi frequency. Second, it requires time-dependent control of two-qubit interactions, since different pulses should act conditioned on whether one of the atoms is de-excited or whether they are both excited. This is necessary in order to counteract unwanted nonadiabatic excitations that can drive the system into entangled states in the $\{\ket{11}, \ket{d_+}\}$ subspace. We here solve the first issue by means of the eCD driving technique~\cite{Petiziol2018}: the imaginary part of the pulses is approximated by introducing additional real Fourier components in the original Hamiltonian, oscillating quickly at a frequency $\omega \gg f_k(t)$. The amplitude and phase of such oscillations is found by enforcing the evolution operator at the end of the oscillation period to match, up to first order in $1/\omega$, the exact counterdiabatic propagator (see Appendix B). In other words, the eCD oscillations approximate the Hamiltonian $H_{\mathrm{CD}}(t)$ with an error scaling with $O(\omega^{-3/2})$ \cite{Petiziol2018}.

A suitable eCD Hamiltonian is found to have the form
\begin{align} 
H_{\mathrm{eCD}}(t) = & \big[g_1(t) (\ket{r}\!\bra{1}\otimes P_0 + P_0 \otimes \ket{r}\!\bra{1}) \nonumber \\
&  + g_2(t) (\ket{r}\!\bra{1}\otimes P_1 + P_1 \otimes \ket{r}\!\bra{1})
+ \mathrm{H.c.} \big]\nonumber \\
& + g_3(t) (P_r \otimes \mathds{1} + \mathds{1}\otimes P_r) ,
\label{eq:Hecd}
\end{align}
with $P_x = \ket{x}\!\bra{x}$, for $x=0,1,r$, and with oscillating functions
\begin{align} 
g_1(t) = & \sqrt{\omega f_0(t)} \sin(\omega t), \quad g_2(t) =  \sqrt{\omega f_1(t)} \cos(\omega t), \nonumber \\
g_3(t) = &-\sqrt{\omega f_1(t)} \sin(\omega t) +\sqrt{\omega f_0(t)}\cos(\omega t) ,
\label{eq:gk}
\end{align}
To investigate the performance of this shortcut-to-adiabatic protocol, we first analyse the final infidelity $\mathcal{I} = 1-|\!\braket{\psi_{\mathrm{fin}}|\psi(t_f)}\!|^2$ produced starting from an initial state $\ket{\psi} = (\ket{00} +\ket{11})/\sqrt{2}$, which should ideally be converted by the CZ gate into the final state $\ket{\psi_{\mathrm{fin}}}=(\ket{00}-\ket{11})/\sqrt{2}$.
This is compared with the original adiabatic protocol in Fig.~\ref{fig:2}, where the infidelity is depicted as a function of the total protocol time and of the Rydberg blockade for the adiabatic $H(t)$ [Fig.~\ref{fig:2}(c)] and shortcut-to-adiabatic $H_{\mathrm{eCD}}(t)$ [Fig.~\ref{fig:2}(d)] protocols. Fidelity regions of interest, above $1-\mathcal{I}=0.999$, require the adiabatic protocol to work in the parameter region of both large blockade and protocol time. When reducing the protocol time for the adiabatic gate, the pulse amplitudes need to be increased in order to maintain the same area. However, this is accompanied by an increase of the phase error mentioned in Sec.~\ref{sec:adiabaticCZ}, causing significant fidelity loss when accelerating the gate with finite blockade. On the contrary, the eCD drive bypasses the generation of the unwanted phase, since it implements only $H_{\mathrm{CD}}(t)$. It is thus able to attain large fidelities for most of the parameter space explored, and in any case outperforms the purely adiabatic scheme. This confirms the efficiency of the modified protocol, on the one hand, and shows also that it is very stable with respect to the choice of timing and blockade parameters. Considering as an example the value of the Rydberg interaction of $V=100$ MHz, a fidelity above 0.998 can be reached in $T_\mathrm{tot}=0.26\;\mu$s, for pulse parameters $\Omega_\mathrm{max}/2\pi=17$ MHz, $\Delta_\mathrm{max}/2\pi=23$ MHz, and a driving frequency $\omega = 300$~MHz. 

\edit{Similarly to the adiabatic protocol, also the modified one necessitates a minimal total protocol time to access a given high-fidelity region. This phenomenon, reminiscent of quantum speed limits~\cite{Deffner2017}, descends indirectly from the increasing violation of adiabaticity for shorter durations. Indeed, as the protocol time is shortened, the exact CD field compensating for nonadiabatic transitions varies on a faster timescale. Specifically, the time derivative of $\partial_t H_{\mathrm{CD}}(t)$ is ruled by $\partial_t f_0(t)$ and $\partial_t f_1(t)$, which scale like $1/T^2_{\mathrm{tot}}$ according to Eq.~\eqref{eq:fk}. As a consequence, a larger driving frequency $\omega$ would be needed to approximate ${H}_{\mathrm{CD}}$ with sufficient precision to maintain a constant fidelity while the protocol time is reduced. Similar effects, in the context of the Landau-Zener-Majorana-St\"uckelberg problem, have been studied, e.g., in Ref.~\cite{Petiziol2018}.}

\edit{To fully characterize the performance of the gate, an experiment could perform a full quantum process tomography (QPT)~\cite{Nielsen2010, Mohseni2008}. According to Kraus' theorem, the qubit gate dynamics can be described as a quantum map $\mathcal{E}(\cdot) = \sum_j K_j(\cdot) K_j^\dagger,$ where the argument is the system's (pure or mixed) state, with Kraus operators $K_j$ satisfying $\sum_j K_j^\dagger K_j=\mathds{1}$. Expanding the operators $K_j$ on tensor products of single-qubit Pauli matrices, $\{\sigma_i\}_{i=0,\ldots,3} = \{\mathds{1}, \sigma_x,\sigma_y,\sigma_z\}$, one obtains
\begin{equation}
\mathcal{E}(\cdot)  = \sum_{i} K_i (\cdot) K_i^\dagger  = \sum_{ij} \chi_{ij} A_i (\cdot) A_j^\dagger,
\end{equation}
which defines the $\chi$-matrix fully characterizing the quantum process~\cite{Mohseni2008}, where $\{A_i\}=\{\sigma_j\otimes \sigma_k\}_{j=0, \dots, 3; k=0,\dots,3}$. Figure~\ref{fig:2}(e) shows the deviation of the ideal $\chi$-matrix from the here proposed eCD-based CZ gate. While simulations include also the Rydberg states, the process tomography shown in Fig.~\ref{fig:2}(e) is computed in the computational subspace of the two qubits. The matrix elements of the eCD $\chi$-matrix differ from the ideal case by less than $10^{-3}$, confirming the effectiveness of the eCD gate for all possible input states.  
The QPT was computed using the QuTip Python library~\cite{Johansson2012, Johansson2013}.}

We see from the structure of the effective Hamiltonian in the first two lines of Eq. \eqref{eq:Hecd} that the required pulse is not separable with respect to the two atoms. This can be overcome by approximating $f_0 \approx f_1$, what would lead to disentangled counterdiabatic pulses that are much easier to implement in practice. In this case the eCD Hamiltonian can be realized as
\begin{align} 
H_{\mathrm{eCD}}'(t) = & \sqrt{\omega \bar{f}}\big[\sin(\omega t) (\ket{r}\!\bra{1}\otimes \mathds{1} + \mathds{1} \otimes \ket{r}\!\bra{1}) + \mathrm{H.c.} \nonumber \\
& + \cos(\omega t) (P_r \otimes \mathds{1} + \mathds{1}\otimes P_r)\big] ,
\label{eq:Hecd_mean}
\end{align}
where we take $\bar{f}$ to be the arithmetic mean between the two original pulses $f_0$ and $f_1$. In contrast to the eCD protocol discussed above, this now approximated protocol has to be combined with the original $H(t)$ from Eq. \eqref{eq:Ht} in order to guarantee the following of the desired state evolution. 

\begin{figure}[t!]
\includegraphics[width=\linewidth]{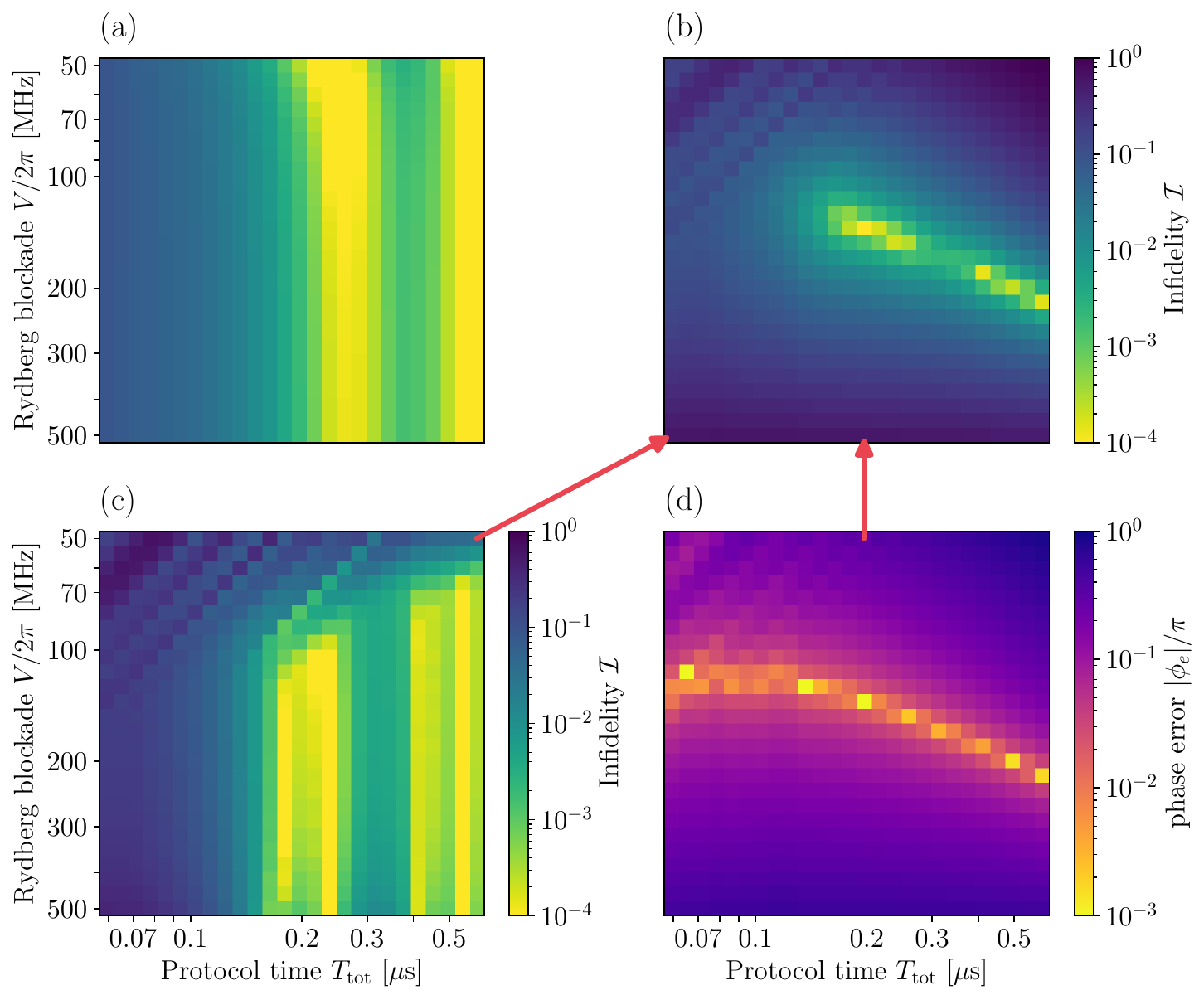}
\caption{Infidelity for the separable accelerated protocol for initial states $(\ket{00}+\ket{01})/\sqrt{2}$ (a), $(\ket{00}+\ket{11})/\sqrt{2}$ (b) and $\ket{11}$ (c) and relative phase error $|\phi_e|/\pi=|1-|\mathrm{arg}\braket{11|\psi_\mathrm{fin}}|/\pi|$ accumulated for initial state $\ket{11}$ (d). The arrows indicate, that the infidelities of the superposition state can be decomposed into a population error, see (c) and a relative phase error of the sensitive state $\ket{11}$, see (d). We use $H(t)$ from \eqref{eq:Ht} with $\Omega_\mathrm{max}/2\pi=17$ MHz and $\Delta_\mathrm{max}/2\pi=23$ MHz, in combination with the approximated eCD method with $\omega=1$ GHz.}
\label{fig:3}
\end{figure}

Results corresponding to Fig. \ref{fig:2}(c,d) are shown in Fig. \ref{fig:3} for initial states involving either $\ket{01}$ or the Rydberg interaction sensitive state $\ket{11}$. The structure along the $x$-axis in the fidelity in (a) and (c) arises from population fluctuations produced by the micromotion of the oscillating Hamiltonian. Similar variations on a smaller scale can already be seen in Fig. \ref{fig:2}(d).  
The results of the superposition state $(\ket{00} + \ket{11})/\sqrt{2}$ (b) are obviously less good than in Fig. \ref{fig:2}.
There is a parameter range, see the maximal fidelity area in Fig. \ref{fig:3}(b), in which the achieved fidelities are nevertheless high. This specific regime was analysed further and we can understand 
the effect by separating the error introduced by the infidelity of state $\ket{11}$, see Fig. \ref{fig:3}(c), and by its unwanted phase error $\phi_e=\pi-|\mathrm{arg}\braket{11|\psi_\mathrm{fin}}|$, see Fig. \ref{fig:3}(d). The fidelity of the superposition state $(\ket{00} + \ket{11})/\sqrt{2}$ in (b) is determined mainly by the phase. This explains the overall lower values and in particular the bright area of higher fidelity. The approximated $H_\mathrm{eCD}$ generates an additional phase which has opposite sign compared to the original phase error \eqref{eq:phaseerror} and decreases for small Rydberg blockades, therefore almost cancelling the total phase error along the bright area. This interesting interference phenomenon allows us to find such regions of high-fidelity.

Having characterized the very promising performance of the eCD gate at the level of a single population transfer, we next analyze the impact of these fidelities in an application to a practical proof-of-concept quantum algorithm.

\section{Application to a QEC code}

\begin{figure}[t!]
\includegraphics[width=\linewidth]{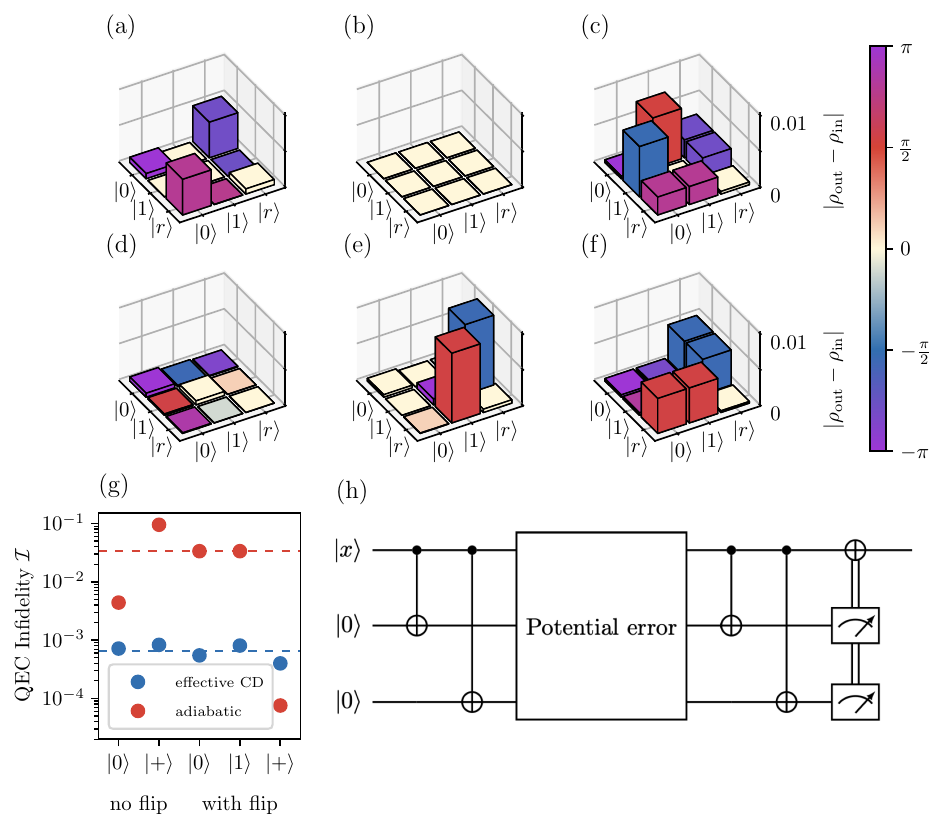}
\caption{Realization of a minimal quantum-error correction algorithm based on the eCD CZ gate with pulses \edit{$\Omega_\mathrm{max}/2\pi=17$ MHz, $\Delta_\mathrm{max}/2\pi=23$ MHz,} protocol time $T_\mathrm{tot}=0.594\;\mu$s, \edit{blockade strength $V/2\pi=500$ MHz} and eCD frequency $\omega=350$ MHz. The quantum circuit is represented in (h), while in (a)-(c) and (d)-(e) the difference of the matrix elements of the recovered density matrix for the corrected qubit is shown. The first and second row corresponds to the case in which no flip or one flip have occurred, respectively, for three different initial states, $\ket{0}$ [(a) and (d)], $\ket{1}$ [(b) and (e)] and $(\ket{0} + \ket{1})/\sqrt{2}$ [(c) and (f)]. The colorscale indicates the phase of the residual matrix elements. \edit{The infidelity of the different configurations is shown in (g) for the circuit based on the eCD gate (blue) compared to the same results using the adiabatic CZ gate in \cite{Saffman2020} (red). We excluded the initial state $\ket{11}$ without a bitflip, as this configuration is dark to the CZ gate. The dashed lines show the average over the five states.} In all cases the minimum fidelities \edit{achieved by the eCD gate} are larger than $0.999$. }
\label{fig:4}
\end{figure}

As an exemplary application of the modified controlled-phase gate, we study a minimal implementation of a quantum-error correcting scheme. We consider the simplified three-qubit code circuit depicted in Fig.~\ref{fig:3}(h)~\cite{Foot2004}, which can correct a potential flip occurring to the first physical qubit only. It involves a sequence of CNOT gates with two ancillary qubits which encodes an input state $\ket{\psi_{\mathrm{in}}}=\alpha\ket{0} + \beta\ket{1}$, prepared in the first qubit, into a superposition $\ket{\psi_{L}}=\alpha\ket{0_L} + \beta\ket{1_L}$ of the logical code words $\ket{0_L}=\ket{000}$ and $\ket{1_L}=\ket{111}$. Error correction is implemented via syndrome measurements on the ancillary qubit, which is preceded by additional CNOT gates, and a conditional recovery~\cite{Foot2004, Nielsen2010}. In the Rydberg realization considered here, the CNOT gates are implemented via single-qubit Hadamard gates and the modified CZ gate, according to \cite{Nielsen2010} 
\begin{equation}
\text{CNOT} = R_c^x(\pi)R_t^y(-\pi/2) \text{CZ} R_t^y(\pi/2)R_c^x(\pi),
\end{equation}
where the labels $c$ and $t$ indicate the control and target qubit, respectively, and $R_{c/t}^i(\theta) = \exp(i \theta \sigma_{c/t}^i /2)$ is a single-qubit rotation. To focus on the impact of CZ gate errors, other operations such as measurement and recovery are assumed to be perfect.  We study the performance of the code by inspecting the difference between the output density matrix post error correction and the input density matrix. This is shown in Fig.~\ref{fig:4}, where the recovery in the case of no error occurrences (first row) is compared with the case of a bit-flip occurrence (lower row), for three different initial states $\ket{0}$ [(a) and (d)], $\ket{1}$ [(b) and (e)] and $(\ket{0} + \ket{1})/\sqrt{2}$ [(c) and (f)]. For all input states, all matrix elements of the reconstructed density matrix match the input $\rho$ up to errors below $10^{-2}$. \edit{In Fig.~\ref{fig:4}(g), we further compare the fidelity results obtained by the eCD gate with those obtained by the adiabatic gate proposed in \cite{Saffman2020}. For almost all error configurations, the eCD method outperforms the adiabatic approach, and ensures very high fidelity of recovery, with similar values for all configurations.}

These results thus confirm that the modified gate enables the implementation of high-fidelity quantum operations, not creating errors that can become unexpectedly amplified during the algorithm. For the case of no error, best recovery is obtained for an initial state $\ket{1}$, which is because the state $\ket{1}$ is dark to the CZ gate after application of the single-qubit Hadamard gate. For the error case, the best recovery corresponds to the input $\ket{0}$ state.

\section{Conclusion}

We have proposed a shortcut-to-adiabatic controlled-phase gate for Rydberg atoms based on effective counterdiabatic fields. The starting protocol from \cite{Saffman2020} has the advantage that it avoids populating additional intermediate states similarly to STIRAP-like protocols \cite{Moller2008, Petiziol2020}. Our proposed accelerated version is less prone to decoherence occurring on longer time scales, while it remains intrinsically stable to its adiabatic backbone. 
\rev{The enhanced performance of the gate is verified in numerical simulations both in single-state transfer fidelities and through full quantum process tomography, confirming the effectiveness of the gate for any choice of input state}.
 Rydberg-interaction induced phase errors are irrelevant if only the counterdiabatic field $H_{\mathrm{CD}}$ is implemented, making the protocol interesting for a wide parameter range, without the need of additional phase compensation \cite{Maller2015}. \rev{We finally exemplify the application of the eCD gate to a practical quantum computation, by discussing its use in a proof-of-concept quantum error correction algorithm and demonstrating typically a large fidelity gain as compared to the purely adiabatic approach.} Future work will be dedicated to disentangle the effective counterdiabatic pulses with respect to the two atoms not only approximately as shown above, possibly following ideas proposed in \cite{Han2016, Su2016}.

\acknowledgments{S.W. acknowledges funding by the National Recovery and Resilience Plan (NRRP), Mission 4 Component 2 Investment 1.3 -- Call for tender No. 341 of 15/03/2022 of Italian Ministry of University and Research funded by the European Union -- NextGenerationEU, Project number PE0000023, Concession Decree No. 1564 of 11/10/2022 adopted by the Italian Ministry of University and Research, CUP D93C22000940001, Project title ``National Quantum Science and Technology Institute'' (NQSTI).} 

\appendix

\section{Adiabatic elimination}
\label{app:a}
In the limit of large Rydberg interaction $V$, the state $\ket{rr}$ is barely populated during the evolution and can thus be adiabatically eliminated as detailed in the following. Then only the states $\ket{rr}$, $\ket{d_+}$ and $\ket{11}$ are coupled to each other, and the time-dependent Schr\"odinger equation thus gives the following system of equations in the manifold spanned by such states,  
\begin{subequations}
\begin{align}
i\frac{\partial_t c_{rr}(t)}{dt} & = [2\Delta(t) + V]c_{rr}(t) +\frac{\Omega(t)}{\sqrt{2}} c_{+}(t),\\
i\frac{\partial_t c_{+}(t)}{dt} & = \frac{\Omega(t)}{\sqrt{2}} [c_{rr}(t) + c_{11}(t)] + \Delta c_{+}(t), \label{eq:ad1}\\
i\frac{\partial_t c_{11}(t)}{dt}& = \frac{\Omega(t)}{\sqrt{2}} c_+(t), \label{eq:ad2}
\end{align}
\end{subequations}
for the corresponding coefficients $c_{rr}(t)$, $c_{+}(t)$ and $c_{11}(t)$ of the wavefunction.
Assuming that $V$ is sufficiently strong for the population of level $\ket{rr}$ not to vary during the dynamics, we impose $\partial_t c_{rr}(t)=0$. This gives 
\begin{equation}
c_{rr}(t) = -\frac{\Omega(t)}{\sqrt{2}[ 2\Delta(t) + V]} c_+(t).
\end{equation}
Inserting this result in the remaining Eqs.~\eqref{eq:ad1} and \eqref{eq:ad2}, and keeping leading-order terms in $V^{-1}$, one obtains the two-state Hamiltonian for $\ket{11}$ and $\ket{d_+}$
\begin{equation}
H = \frac{1}{2}\begin{bmatrix}
0 & \sqrt{2}\Omega(t) \\
\sqrt{2}\Omega(t) & 2\Delta(t),
\end{bmatrix}
\end{equation}
as reported in Eq.~\eqref{eq:Hk}.

\section{Derivation of the eCD field}
\label{app:b}

The idea of the eCD scheme is to construct a Hamiltonian $H_{\mathrm{eCD}}(t)$ which contains fast oscillations in controllable parameters, whose `averaged' effect is the realization of $H_{\mathrm{CD}}(t)$. Namely, one forces the propagator produced by $H_{\mathrm{eCD}}(t)$ to approximately match the propagator generated by $H_{\mathrm{CD}}(t)$. This procedure can be understood as Floquet engineering~\cite{Bukov2015, Eckardt2017, Goldman2014} of a time-dependent Hamiltonian, and is typically done by computing first terms of a high-frequency Magnus expansion~\cite{Petiziol2018}. 
We show now that the proposed eCD field of Eq.~\eqref{eq:Hecd} generates the counterdiabatic dynamics given by the Hamiltoninan of Eqs.~\eqref{eq:Hcd01} up to order $1/\omega$.

Starting from Eq.~\eqref{eq:Hecd}, the first terms $H_{\mathrm{eff}}^{(n)}$ of the efffective Hamiltonian computed via the Magnus expansion~\cite{Blanes2009} generated by $H_{\mathrm{eCD}}(t)$ at the end of the oscillation period $T=2\pi/\omega$ read
\begin{align}
 H_{\mathrm{eff}}^{(0)}  =& \int_0^T H_{\mathrm{eCD}}(t) dt = 0,\\
H_{\mathrm{eff}}^{(1)} = -&\frac{1}{2}\int_0^T dt_1\int_0^{t_1}dt_2[H_{\mathrm{eCD}}(t_1), H_{\mathrm{eCD}}(t_2)]  \\
= & \frac{f_0(t)}{2} \big[(-i\ket{r}\!\bra{1}+ i \ket{1}\!\bra{r}) \otimes P_0 \\
&\qquad  + P_0 \otimes (-i\ket{r}\!\bra{1} + i \ket{1}\!\bra{r})\big] \\
& +\frac{f_1(t)}{2} \big[-i \ket{d_+}\!\bra{11} + i\ket{11}\!\bra{d_+}\big].
\end{align}
The oscillatory $H_{\mathrm{eCD}}(t)$ thus produces an effective end-of-period Hamiltonian with the same desired structure of $H_{\mathrm{CD}}(t)$.

\end{document}